\documentclass[12pt, draftclsnofoot, onecolumn]{IEEEtran}	% regular single-column
 \usepackage{amsmath,amssymb}
 \usepackage{subfigure}
 \usepackage{graphicx,graphics,color,psfrag}
 \usepackage{cite,balance}
 \usepackage{algorithm}
 \usepackage{accents}
 \usepackage{amsthm}
 \usepackage{bm}
 \usepackage{url}
 \usepackage{algorithmic}
 \usepackage[english]{babel}
 \usepackage{multirow}
 \usepackage{enumerate}
 \usepackage{cases}
 \usepackage{stfloats}
 \usepackage{dsfont}
 \usepackage{color,soul}
 \usepackage{amsfonts}
 \usepackage{cite,graphicx,amsmath,amssymb}
 \usepackage{subfigure}
 \usepackage{fancyhdr}
 \usepackage{hhline}
 \usepackage{graphicx,graphics}
 \usepackage{array,color}
 \usepackage{amsmath}
 \usepackage{amsthm}
 \usepackage{framed} 
 \usepackage{multirow}
\usepackage{hyperref}
\usepackage{setspace}

\def\({\left(}
\def\){\right)}

\def\b0{{\mathbf{0}}}

\def\papertitle{ \huge Over-the-Air Computing for Wireless Data Aggregation in Massive IoT}

\IEEEoverridecommandlockouts 

\begin{document}

\title{ \fontsize{21}{21}\selectfont \papertitle}
\author{Guangxu Zhu, Jie Xu, Kaibin Huang, and Shuguang Cui
\thanks{ G.~Zhu is with the Shenzhen Research Institute of Big Data, Shenzhen, China. He was formerly with The University of Hong Kong. J. Xu and S. Cui are with FNii and SSE at the Chinese University of Hong Kong (Shenzhen), Shenzhen, China. K. Huang is with EEE at the University of Hong Kong, Hong Kong. 
%J. Zhang is with the Dept. of EIE at the Hong Kong Polytechnic University. Email: jun-eie.zhang@polyu.edu.hk. (Corresponding author: K. Huang).
}
%\thanks{J. Zhang is with the Dept. of ECE at the Hong Kong University of Science and Technology. Email: eejzhang@ust.hk. }
\vspace{-10mm}
}
\maketitle

%\thispagestyle{fancy} 
%\fancyhead{}
% \lhead{\small The Reply Letters to the Reviewers' Comments can be Downloaded via 
%{\bf \url{https://www.eee.hku.hk/~wirelesslab/resources/Reply.pdf} }
%} 
 %?????
%\chead{}
%\rhead{}
%\lfoot{}
%\cfoot{\thepage} %current page number
%\rfoot{}
\renewcommand{\headrulewidth}{0pt} %?????????????????

\vspace{-4mm}
\begin{abstract}
\emph{Wireless data aggregation} (WDA), referring to aggregating data distributed at devices (e.g., sensors and smartphone), is a common operation in 5G-and-beyond machine-type communications to support  \emph{Internet-of-Things} (IoT), which lays the foundation for diversified applications such as distributed sensing, learning, and control. Conventional WDA techniques that are designed based on a separated-communication-and-computation principle encounter difficulty in accommodating the massive  access under the limited radio resource and stringent latency constraints imposed by emerging applications (e.g, auto-driving).
%due to the limited radio resource constraint. 
To address this issue, \emph{over-the-air computation} (AirComp) is being developed as a new WDA solution by seamlessly integrating computation and communication. By exploiting the waveform superposition property of a multiple-access channel, AirComp turns the air into a computer for computing and communicating functions of distributed data at many devices, thereby allowing low-latency WDA over massive devices. In view of growing interests on AirComp, this article provides a timely overview of the technology by introducing basic principles, discussing advanced techniques and applications, and identifying promising research opportunities.

 \end{abstract}

%\begin{IEEEkeywords}
%Over-the-air computation, multiple-input multiple-output, beamforming, channel feedback
%\end{IEEEkeywords}

\IEEEpeerreviewmaketitle

\vspace{-3mm}
\section{Introduction}

%Sensing, computation, and communication are three basic functionalities of B5G wireless networks [212]. Traditionally, these three functionalities have been carried out independently. Hence, it is necessary to allocate wireless resources for each functionality, resulting in a high resource consumption. In the case of massive access, the resources required to support these functionalities might be prohibitive. Hence, it is desirable to jointly design these three functionalities to improve the efficiency of massive access. For instance, the transmission of sensed signals over multiple access channels can also be exploited to perform computations by applying over-the- air computation techniques in [213]. In this scenario, the limited wireless resources can be utilized with high efficiency, especially for massive access. Therefore, the convergence of sensing, computation, and communication is an important future direction for cellular IoT in B5G wireless networks.

Driven by the vision of \emph{Internet-of-Things} (IoT), which is expected to revolutionize the way we live and work through providing ubiquitous connectivity to everything, the evolution of wireless communications from 1G to 5G has witnessed a paradigm shift from human-type communications towards machine-type communications. Particularly, it is predicted by the \emph{Global System for Mobile Communications Association} (GSMA) that the number of IoT devices will reach 75 billion by 2025, which is much larger than that of mobile phone users.   
To provide wireless connectivity to such a gigantic number of devices poses a grand challenge to the existing wireless systems: the scaled-up radio resources required by massive connectivity overwhelm the capacity of the existing systems. While there have been several prior works exploiting emerging techniques such as millimeter-wave communications and massive \emph{multiple-input multiple-output} (MIMO) (see, e.g., \cite{liu2018sparse,chen2020massive} and the references therein) to  support the massive connectivity, such a research gap still cannot be fully closed due to the ever-increasing wireless devices. 
This thus has prompted an increasing number of researchers to depart from the traditional design principle that isolates communication from the subsequent applications, and explore the designs of application-specific wireless technologies integrating  disciplines such as machine learning, computing, and communications \cite{letaief2019roadmap}.

 \begin{figure*}[tt]
\centering
\includegraphics[width=0.85\textwidth]{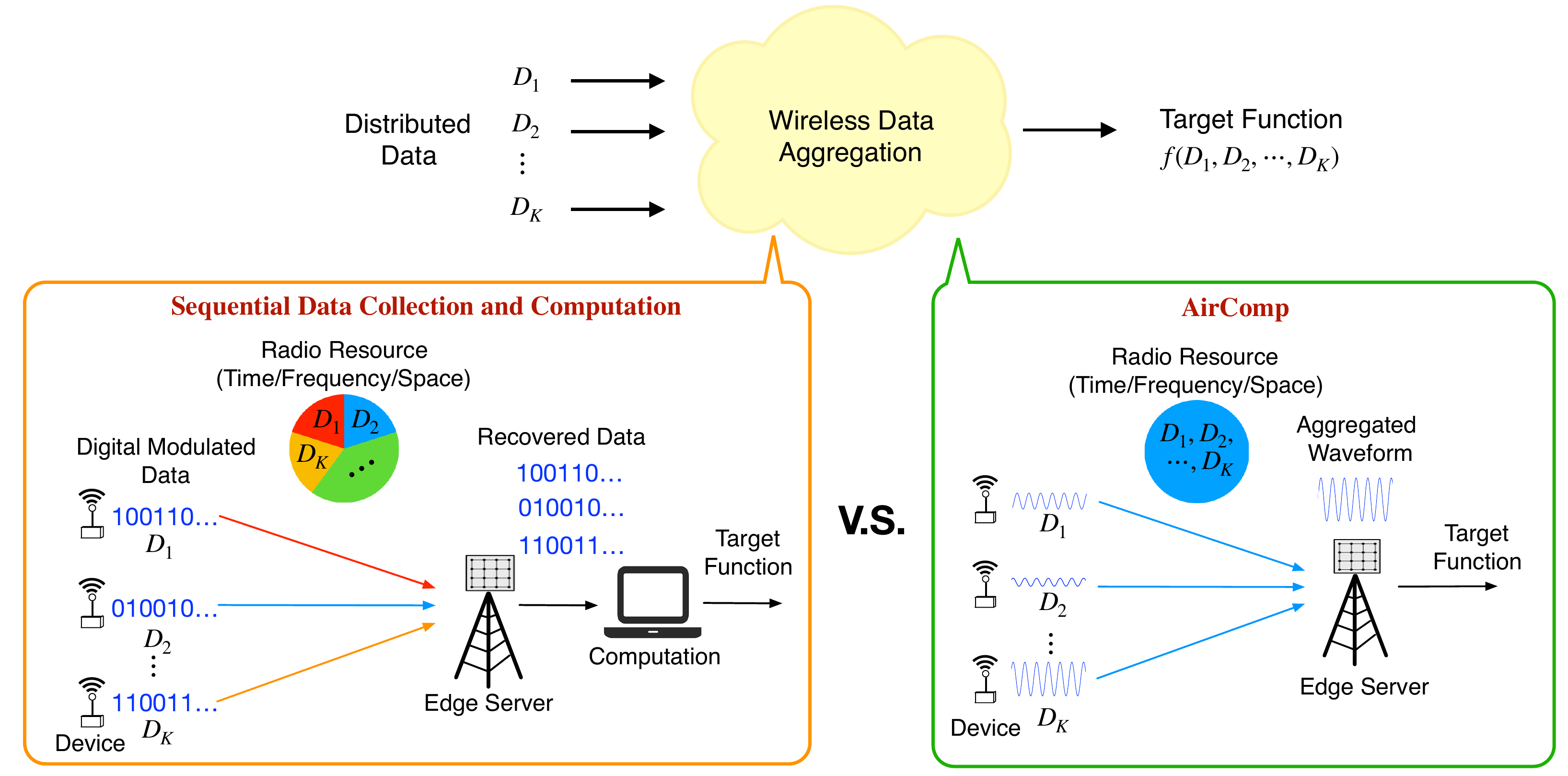}
\caption{Two paradigms for WDA over a MAC: Sequential data collection and computation versus AirComp.}
\label{model1:subfig}
\vspace{-4mm}
\end{figure*}

Aligned with this direction,  a specific class of IoT applications has emerged, which requires an edge server (which can be a base station) to aggregate data distributed at devices with wireless connectivity, termed \emph{wireless data aggregation} (WDA). Such applications include vehicle platooning, drone swarm control, distributed sensing and learning. In such applications, a server is interested in computing a function of distributed data generated by devices. Such data may include \emph{artificial intelligence} (AI) model updates in distributed learning, accelerations and velocities in vehicle platooning or drone swam, and temperature/humidity/chemical-levels in sensing. The applications are either data intensive (e.g., distributed learning) or latency critical (e.g., vehicle platooning). The requirements have motivated researchers to develop a new technology, called \emph{over-the-air computation} (AirComp), to enable efficient WDA over many devices. The basic principle of AirComp is to exploit the waveform superposition property of a wireless channel to realize over-the-air aggregation of data simultaneously transmitted by devices. Simultaneous transmission in AirComp allows each device to access all radio resources instead of only a fraction of them as in the conventional orthogonal multiple access schemes (see Fig. \ref{model1:subfig}), thus allowing high spectrum-efficiency WDA. A vivid interpretation of the key feature of AirComp is to harness interference to help functional computation \cite{GastparTIT2007}, thereby turning the air into a computer.

In the 5G-and-beyond era, we shall see the merging of sensing, computation, and communication in IoT networks. This ongoing trend has driven the rapid development of AirComp to bring it closer to reality.   
As attempts to tackle the practical challenges faced in materializing the promising gain of AirComp, recent research in this field focuses on developing a versatile WDA technology by advancements in different directions including power control, spatial multiplexing, channel feedback, and multi-cell cooperation. In view of growing interests on AirComp, this article provides a comprehensive introduction of the new technology covering the basic principles, advanced techniques, applications, and research opportunities.

\section{AirComp Fundamentals}\label{sec:BP}
%\subsection{Uncoded AirComp}
For ease of exposition, the fundamentals of AirComp are described under single-antenna setting in this section. As mentioned, the basic idea of AirComp is to exploit the analog-wave superposition property of a \emph{multiple-access channel} (MAC). As a result, the signals simultaneously transmitted by synchronized devices are added over-the-air and arrive at the receiver as weighted sum, called the \emph{aggregated signal}, with weights being the channel coefficients. As shown in Fig. 2, the two essential operations for  AirComp are \emph{linear-analog modulation} and \emph{channel pre-compensation} at each transmitter. The former modulates the data values into the magnitudes of the carrier signals; the latter compensates for heterogeneous channel fading of different links.  As a result, each component part of received signal is the transmitted data scaled by a pre-determined factor. Setting the factor uniform for all signals, called \emph{magnitude alignment}, reduces the aggregated signal to the desired average of transmitted distributed data, realizing the AirComp of an average function. 
Essentially, AirComp can be understood as a joint source and channel design in contrast to the classic separation-based design featuring sequential communication and computation. Particularly, it was shown in the landmark work \cite{GastparTIT2007} that AirComp is optimal in terms of minimizing the \emph{mean squared error} (MSE) distortion for the scenario of Gaussian MAC with independent Gaussian sources.  

 %from the rate-distortion tradeoff perspective
  %As noted, the popularity of AirComp is attributed to its simplicity 

%xxx: A surprising answer was provided in the landmark work in [3] from the information- theoretic perspective that coding is unnecessary for AirComp over a Gaussian MAC with independent Gaussian sources. In other words, uncoded AirComp is optimal in this case.

With appropriate data pre/post-processing, the capability of AirComp can go beyond averaging to compute a class of so-called \emph{nomographic functions}, which can generally be expressed as a post-processed summation of multiple pre-processed data-values (see \cite{zhu2018mimo} for the precise mathematical definition). Typical functions in this class include arithmetic mean, weighted sum, geometric mean, polynomial, and Euclidean norm. For example, to compute the geometric mean, the pre-processing is a logarithm function and post-processing an exponential function as presented in \cite{zhu2018mimo}. Interestingly, it has been proven in \cite{buck1976approximate} that any function can be decomposed as a summation form of nomographic functions, indicating that any function can be computed via AirComp in general.

 \begin{figure*}[tt]
\centering
\includegraphics[width=0.8\textwidth]{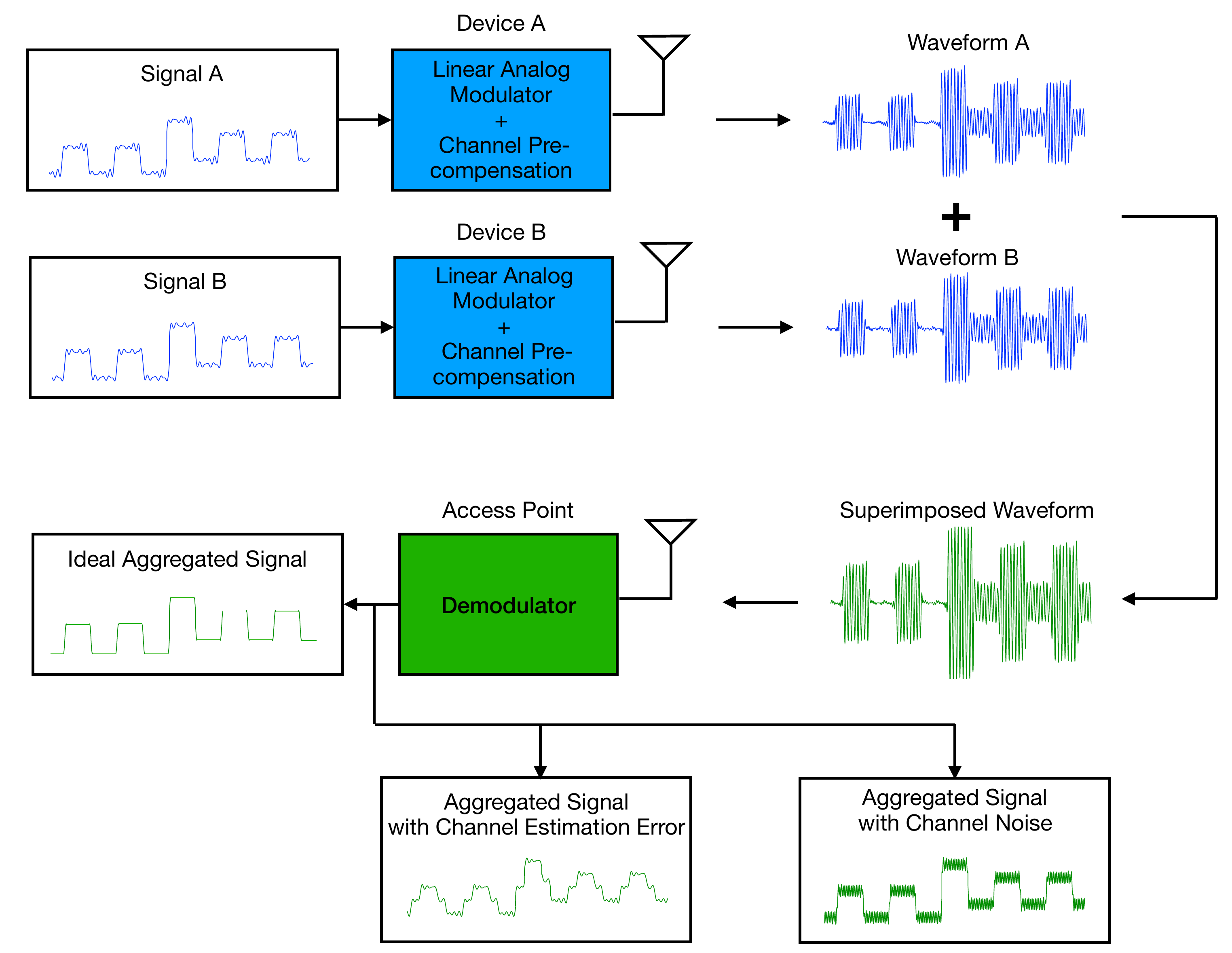}
\caption{ Illustration of basic principle for AirComp.}
\label{fig:AirComp_ill}
\vspace{-4mm}
\end{figure*}

Strict time synchronization in devices' transmissions poses a key challenge for AirComp implementation, but can be overcome using the rich set of existing synchronization techniques. For instance, uplink synchronization in 4G \emph{Long Term Evolution} (LTE) systems relies of a so-called ``timing advance'' mechanism, which can be used to facilitate the AirComp in practice. Specifically, each device estimates the propagation delay and then transmits ahead of time (with a negative time offset equal to the delay) so that the signal always arrives at the base station within the allocated time slot regardless of the device's location. The synchronization accuracy is proportional to the bandwidth of the synchronization channel used for propagation delay estimation. For example, a typical bandwidth of $1$ MHz reins in the timing offset/error to be within $0.1$ microsecond \cite{arunabha2010fundamentals}. Consider the implementation of AirComp in a popular OFDM system, the timing offset simply introduces a phase shift to the received symbol if the offset is shorter than the \emph{cyclic prefix} (CP), which can thus be compensated by sub-channel equalization. The typical CP length in LTE systems is $5$ microseconds which is far longer than the typical timing offset, i.e., $0.1$ microsecond \cite{arunabha2010fundamentals}. Thus the time synchronization for AirComp is  feasible. 

Due to the employed analog modulation, AirComp is exposed to signal distortion caused by channel fading, noise, and channel estimation error as illustrated in Fig. \ref{fig:AirComp_ill}. The distortion of the received functional values can be suitably measured using the MSE with respect to the noiseless ground truth, which is a commonly used performance metric for AirComp and termed the computation error.

It is worth noting that AirComp is fundamentally different from the concept of (uplink) \emph{non-orthogonal multiple access} (NOMA), in various aspects such as objectives and performance metrics. In terms of objective,  AirComp aims at computing a certain function of the distributed data at devices by \emph{harnessing} inter-user ``interference''. In contrast, NOMA aims at decoding individual data streams from the simultaneously transmitted signals by successively \emph{canceling} the harmful inter-user interference. In terms of performance metric, AirComp concerns the accuracy of distributed function computation with guaranteed computation rate, which refers to the number of received functional values per channel use. In contrast, NOMA concerns the classic communication rate metric under  guaranteed  decoding reliability.

\section{Advanced AirComp Techniques} \label{sec:techniques}
%A set of advanced AirComp techniques need to be 

Recent research on AirComp focuses on advanced techniques aiming at boosting the computation rates, reducing computation errors, or supporting large-scale deployment. 

%cope with the computation distortion due to channel noise, fading, and hardware constraints,
%Materializing the promising gain of AirComp requires the development of a set of advanced AirComp techniques ranging from power control to interference management, to cope with the computation distortion due to channel noise and fading, and hardware constraints. In this section, several enabling techniques for practical AirComp implementation are introduced as follows.
% with emphasis on the principles, key challenges, and research opportunities.

%\vspace{-3mm}
\subsection{Power Control for AirComp}
In AirComp, channel inversion is usually implemented at the transmitters by adjusting their transmission power to achieve magnitude alignment at the receiver. However, when one or more individual channels are in deep fade, enforcing the magnitude-alignment constraint can result in large AirComp errors. The reason is that a very small alignment factor has to be chosen to make it possible for all devices including those with weak links to perform channel inversion, which weakens the aggregated signal and hence amplifies the negative effect of channel noise. This suggests that uniform channel inversion may not be always desirable and the optimal power-control policy for AirComp should be adapted to multiuser channel states. Recently, it was shown in \cite{cao2019optimal} that the optimal policy for the case of independent sources exhibits a ``binary'' structure. Specifically, devices with weak channel gains, which are below a derived threshold, should transmit with full power while others should perform channel inversion. The departure of such power-control policy from the optimal ``water-filling'' one in conventional  communication systems highlights the difference between WDA and sum-rate maximization \cite{cao2019optimal}.

{ \bf \underline{Some research opportunities:}}

%Some research opportunities in this direction are discussed as follows:
\begin{itemize}
\item {\bf Optimal power control for AirComp with correlated sources}: 
Source correlation usually exists in practical applications such as distributed sensing. In this case, the optimal power control for AirComp remains unknown while the discussed binary strategy from  \cite{cao2019optimal} is optimal only in the case without source correlation. The challenge lies in that the source correlation makes the power control at different devices highly coupled in the MSE objective, rendering a much sophisticated optimization problem.

\item {\bf Robust power control for AirComp}: 
In practice, inaccuracy usually exists in the channel estimation process. Power control using imperfect \emph{channel-state information} (CSI) can lead to large AirComp errors. Therefore, it is important to characterize the effect of imperfect CSI on the AirComp performance. Leveraging the results, robust power control algorithms can be designed to optimize the worst-case AirComp performance. 

\end{itemize}

%\begin{figure}[tt]
%\centering
%\includegraphics[width=0.4\textwidth]{Figures/Power_control_illustration.pdf}
%\caption{ Illustration of the optimal power control policy for  AirComp.}
%\label{fig:power_contl_ill}
%\vspace{-3mm}
%\end{figure}

%\begin{itemize}
%
%\item Difference between power control for AirComp network and conventional one for rate-centric communication networks.
%
%\end{itemize}

\subsection{MIMO AirComp}

\begin{figure}[tt]
\centering
\includegraphics[width=0.5\textwidth]{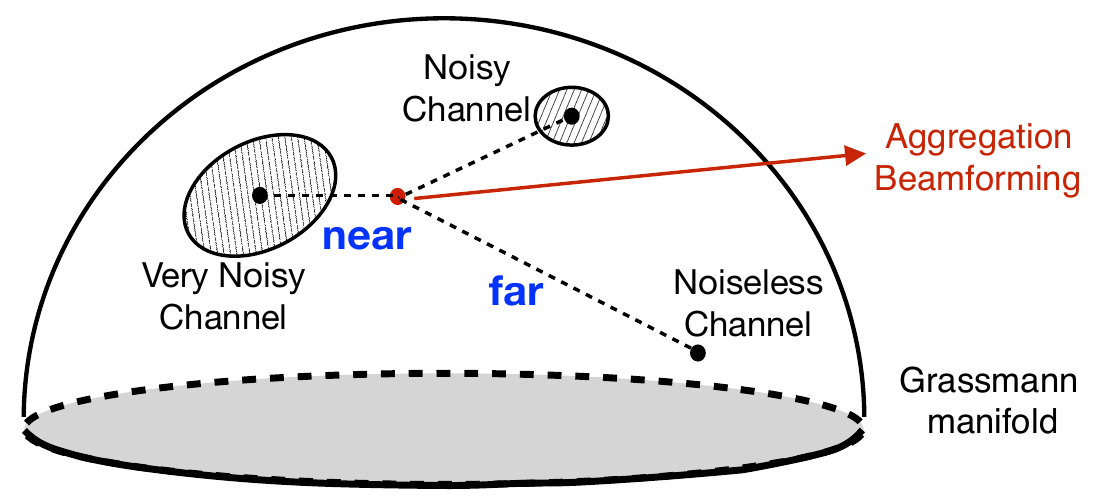}
\caption{ \!\!\!Geometric \! interpretation \!of \! aggregation \! beamforming \! for  AirComp\!.}
\label{fig:MIMO_AirComp_ill}
\vspace{-3mm}
\end{figure}

Some emerging WDA applications are either latency sensitive  or data intensive. To support such applications motivates the high-rate AirComp by spatial multiplexing over MIMO channels, or equivalently the realization of vector-function AirComp. MIMO AirComp differs from its single-antenna counterpart in two ways. First, the channel-inversion power control of the latter is replaced with zero-forcing precoding. Second, the multi-antenna server attempts to apply receive beamforming, called \emph{aggregation beamforming}, to achieve \emph{simultaneous magnitude alignment} (or simultaneous aggregation) of spatially multiplexed multiuser signals so as to receive parallel functional streams; such an operation is unavailable for a single-antenna server. One key challenge on designing MIMO AirComp is to optimize the aggregation beamformer for minimizing the MSE of vector-function AirComp. One approximate solution was obtained in \cite{zhu2018mimo} which is presented on a \emph{Grassmann manifold} where the subspace corresponding to a MIMO channel matrix is mapped to a singe point and so is the aggregation beamformer. By approximate MSE minimization, the beamformer is designed as the weighted sum of individual MIMO channel subspaces with the weights determined by the channel strengths. As illustrated in Fig. \ref{fig:MIMO_AirComp_ill}, the geometric interpretation is that the optimal aggregation beamformer tends to align closer with relatively noisy MIMO channels and less with the less noisy ones, so as to equalize the noise levels in different channels to achieve overall AirComp error reduction.

{\bf \underline{Some research opportunities}:}

\begin{itemize}

\item {\bf Optimization of precoders and aggregation beamformer}. 
The optimal design for aggregation beamforming remains unknown. As channel-inversion power control is sub-optimal for single-antenna AirComp, zero-forcing precoding is sub-optimal for MIMO AirComp. To minimize the MSE, it is desirable to jointly optimize the precoders at devices and the aggregation beamformer at the server, which appears to be a challenging problem to solve.

\item {\bf Diversity-multiplexing tradeoff for MIMO AirComp}.
For conventional MIMO communications systems, there exists a fundamental diversity-multiplexing tradeoff in terms of reliability and sum-rate performance. A similar tradeoff also holds for MIMO AirComp where the spatial \emph{degrees-of-freedom} can be applied either to spatially multiplex functional streams or to reduce their errors. Quantifying the tradeoff helps the understanding of the fundamental limit of MIMO AirComp. 

\end{itemize}

\vspace{-3mm}
\subsection{Multi-cell AirComp}
In next-generation IoT, the relevance of AirComp to different types of applications and its being a promising low-latency solution suggest the need of considering its large-scale deployment in a multi-cell network. Multi-cell AirComp can be implemented in two modes, namely hierarchical AirComp and coordinated AirComp. In hierarchical AirComp, a centralized server aggregates AirComp results output by local servers through backhaul links to scale up the aggregation gain, e.g., training a larger model or exploiting a larger dataset in the context of distributed edge learning. On the other hand, coordinated AirComp aims at supporting coexisting WDA tasks in different cells, each of which is characterized by its application, data type, and target function. The coexisting tasks in coordinated AirComp are exposed to inter-cell interference. This calls for interference management by multi-cell coordination to balance the errors in the coexisting tasks. While multi-cell AirComp is an open area, an initial attempt has been made in \cite{Cao2020cooperative} on understanding the performance limit of coordinated AirComp by quantifying the Pareto boundary of the multi-cell MSE region. 

{\bf \underline{Some research opportunities}:}

\begin{itemize}

\item {\bf Hierarchical AirComp with limited cooperation overhead}:
 The centralized control of the operations at a large number of nodes can incur excessive signaling overhead (e.g., CSI feedback) especially in the case of MIMO channels. To reduce the overhead, it is highly desired to implement the network-wise cooperation in a distributed manner, in which most of the processing can be done locally, requiring only limited
communication between nearby devices. 
%This is an important but yet addressed problem.

\item {\bf Performance and techniques for coordinated MIMO AirComp}: 
The Pareto boundary of the multi-cell MSE region in this scenario is much more sophisticated than that for single-antenna network studied in \cite{Cao2020cooperative}. Specifically, the complexity arises from the need of designing interference management techniques by the joint optimization of precoders, aggregation beamformers, and power control across the whole network.

\end{itemize}

%\begin{itemize}
%\item Existing works on AirComp all assumed a single-cell system with a single function computed within the cell. Extension of AirComp to the multi-cell system is a largely uncharted area. New techniques for inter-cell interference management need be designed to maximize the number of aggregated signal streams in the system. In particular, the classic theory of cooperative power control and  interference-alignment  can be redeveloped to account for AirComp operations and the new performance metrics. %such as signal alignment in magnitude and the learning performance metric including convergence rate and accuracy.
%
%\item Cooperative power control 
%
%\item Interference management
%
%\end{itemize}

\subsection{AirComp with Digital Modulation}

The basic analog AirComp design introduced in Section \ref{sec:BP} requires linear analog modulation, which implicitly assumes that the transmitter can modulate the carrier waveform as desired, freely setting the waveform magnitude as arbitrary real number. However, digital modulation such as \emph{quadratic amplitude modulation} (QAM) is widely used in practical systems such as LTE and 5G, and most existing devices come with embedded digital modulation chips that cannot support an arbitrary modulation scheme.
%The basic AirComp design introduced in Section \ref{sec:BP}  requires linear analog modulation. However, digital modulation such as \emph{quadratic amplitude modulation} (QAM) is widely adopted in practical systems such as LTE and 5G. Consequently, the existing baseband chipsets may not support analog modulation required by AirComp. 
To address this issue, an idea, called \emph{digital AirComp}, was proposed in \cite{zhu2020one} that a  QAM modulator can be treated as a quadratic linear analog modulator with complex amplitude quantization. The idea allows AirComp to be implemented on popular transceiver architectures such as an OFDM transceiver as demonstrated in an application to distributed edge learning in \cite{zhu2020one}.  The digital solution therein features one-bit gradient quantization at devices and a majority-vote based gradient-decoding at the server. Compared with its analog counterpart, the digital AirComp solution enjoys the robustness against noise in function-value detection at a cost of reduced computation resolution to some extent. Surprisingly,  it is experimentally shown in \cite{zhu2020one} that the quantization error in the aggregated gradient does not incur significant loss in learning performance, demonstrating the effectiveness of digital AirComp.

%The basic analog AirComp design introduced in Section \ref{sec:BP} requires linear analog modulation, which implicitly assumes that the transmitter can modulate the carrier waveform as desired, freely choosing the I/Q coefficients as arbitrary real number. However, digital modulation is widely used in practical systems, and most existing devices come with embedded digital modulation chips that cannot support an arbitrary modulation scheme. To facilitate the wide-deployment of AirComp in existing systems without significant changes in hardwares, AirComp should be made compatible with the digital modulation, motivating the development of digital AirComp. Notice that the popular digital \emph{quadratic amplitude modulation} (QAM), is a type of amplitude modulation (albeit with discretization errors) like linear analog modulation. This opens up the possibility to implement digital AirComp based on QAM signals. An initial attempt has been made in  \cite{zhu2020one} for the application of distributed edge learning. The digital solution therein features one-bit gradient quantization at devices and majority-vote based gradient-decoding at the server, which is experimentally shown to achieve comparable performance as its analog counterpart in terms of learning accuracy \cite{zhu2020one}, demonstrating the promise of digital AirComp. 

{\bf \underline{Some research opportunities}:}

\begin{itemize}
\item {\bf Digital AirComp with adaptive modulation}: 
In digital AirComp, the modulation order (or quantization level) serves as a control variable regulating the tradeoff between the functional-value resolution, modulation complexity, and robustness against channel noise. Existing design in [11] assumed a fixed modulation scheme for simplicity.
Optimally adapting the modulation order to the channel state and a latency requirement can further enhance the performance of digital AirComp.

%In digital AirComp, the modulation order (or quantization level) serves as a control variable regulating the tradeoff between the computation resolution and noise-induced error. Existing design in \cite{zhu2020one} assumed a fixed modulation scheme for simplicity. It thus leaves room to enhance digital AirComp with adaptive modulation.

\item {\bf Effect of quantization distortion}: 
The effect of quantization error introduced by digital AirComp on the ultimate computation performance  has not been theoretically understood. A rigorous analysis of such effects for different applications helps further optimization of the quantization scheme in the context of AirComp.
\end{itemize}

\begin{figure*}[tt]
\centering
  \includegraphics[width=0.85\textwidth]{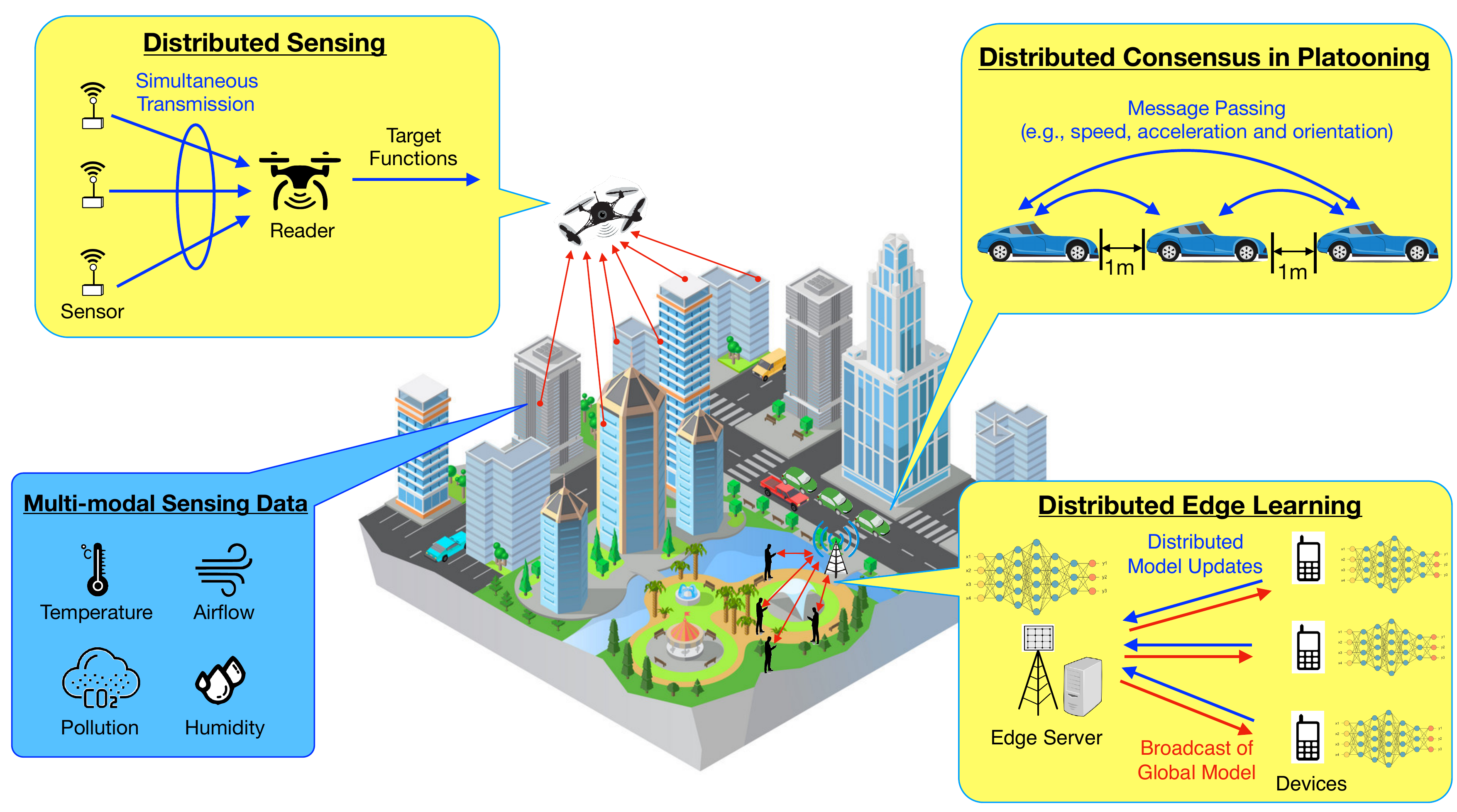}
\caption{Illustration on AirComp applications.}
\label{fig:applications}
\vspace{-4mm}
\end{figure*}

\vspace{-3mm}
\subsection{CSI Feedback for AirComp}
The server requires the CSI of individual uplink channels in order to deploy the aforementioned   AirComp techniques such as power control, aggregation beamforming, and interference management. The usual approach of acquiring the CSI is to let the server sequentially estimate individual channels or let devices feed back their CSI when channel reciprocity is available. However, this may cause excessive latency and overhead when there are many devices. A more intelligent approach is to apply AirComp to accelerate the CSI-feedback process. In other words, AirComp is applied not only in WDA but also in acquiring CSI needed for WDA. This approach was demonstrated in \cite{zhu2018mimo} for enabling aggregation beamforming for MIMO AirComp. Assume the availability of channel reciprocity that allows devices to have the CSI of their individual uplink channels. The feedback design in \cite{zhu2018mimo} involves each device transmits an analog modulated signal computed from its individual CSI so that an over-the-air aggregated signal received by the server can be used to directly extract the desired aggregation beamformer. Note that the design bypasses CSI acquisition to directly receive the beamformer by simultaneous ``one-shot''  transmissions from devices. Consequently, the CSI-acquisition overhead is independent of the number of devices. 

{\bf \underline{Some research opportunities}:}

\begin{itemize}
\item {\bf Robust CSI feedback}: A key drawback of the discussed ``one shot'' feedback scheme is that the acquired aggregation beamformer is exposed to the perturbation by channel noise and interference. This can severely degrade the AirComp performance and calls for robust feedback design. A possible solution is to provide feedback protection by scrambling analog feedback signals or even apply a coding techniques customized for AirComp (e.g., lattice coding \cite{GastparTIT2007}).

\item {\bf CSI-free AirComp}: 
%An alternative solution to tackle the channel feedback overhead in large-scale AirComp networks is to develop blind AirComp technique that exempts the need to acquire CSI. This can be made possible by adopting the randomized encoding scheme at devices that embeds the source data into the distribution of transmitted signals, such that the desired function value of the source data is embedded into the distribution of the aggregated signal, which can be retrieved by (distribution) parameter estimation at server using tools from statistical inference by treating the unknown channel coefficients as hidden variables.
An alternative solution to avoid CSI feedback is to develop CSI-free AirComp techniques. One possible idea is to do randomized encoding at devices that embeds the source data into the distribution of transmitted signals such that the desired function of the distributed data is embedded into the distribution of the aggregated signal. The function can then be retrieved by distribution-parameter estimation at server using tools from \emph{statistical inference}. 
%by treating the unknown channel coefficients as hidden variables.

%An alternative solution to tackle the channel feedback overhead in large-scale AirComp networks is to develop blind AirComp technique  that exempts the need to acquire CSI. This can be made possible by adopting the \emph{randomized encoding} scheme at  devices that embeds the source data into the distribution of transmitted signals, such that the desired function value of the source data is embedded into the distribution of the aggregated signal, which can be retrieved by (distribution) parameter estimation at server using tools from \emph{statistical inference} by treating the unknown channel coefficients as hidden variables. 
%\cite{zhu2018inference}.
 
%How the quantization error contribute to the AirComp error in general, and what is the optimal quantization strategy.
\end{itemize}

\vspace{-3mm}
\section{Applications -- Sensing, Learning, Consensus}\label{sec:app}
AirComp is envisioned to have a wide range of IoT applications related to the areas of distributed sensing, learning, and consensus, as illustrated in Fig. \ref{fig:applications} and discussed in the following.

\vspace{-3mm}
\subsection{Distributed Sensing}
As perhaps the most fundamental operation in IoT networks, distributed sensing is the foundation of many other upper-layer IoT applications such as environmental monitoring and smart city, in which distributed sensing data from a large number of sensors needed to be aggregated and then used for inference and making decisions on how to control the physical environment via actuators. One challenge faced by next-generation distributed sensing is fast WDA in scenarios with high mobility and intensive data uploading. As illustrated in Fig. \ref{fig:applications}, a fusion center mounted on an \emph{unmanned aerial vehicle} (UAV) or ground vehicle can be deployed to monitor a wild environment or a city by collecting environmental sensing data  (e.g., airflow, temperature, pollution, and humidity). Due to the short center-sensor contact and the potential high density of sensors, the conventional ``separated-communication-and-computation'' approach may not be able to meet the stringent latency requirement for fast WDA.

The problem can be solved using AirComp. In many IoT applications, a fusion center is interested in knowing a specific function of distributed data instead of individual data samples. For example, for environmental monitoring, the sensor network should monitor  the average value of noisy temperature readings measured by sensors distributed over a particular area. As another example, in a disaster avoidance system, the interested functional value is the maximum chemical level or temperature over the sensor readings. The integrated functional computation and wireless transmission enables AirComp to support fast WDA in large-scale distributed sensing with multi-access latency independent of the network scale.

\setlength{\textfloatsep}{4pt}
\begin{figure*}[tt]
\centering
\subfigure[Test accuracy ]{\label{Fig:test_acc}\includegraphics[width=0.45\textwidth]{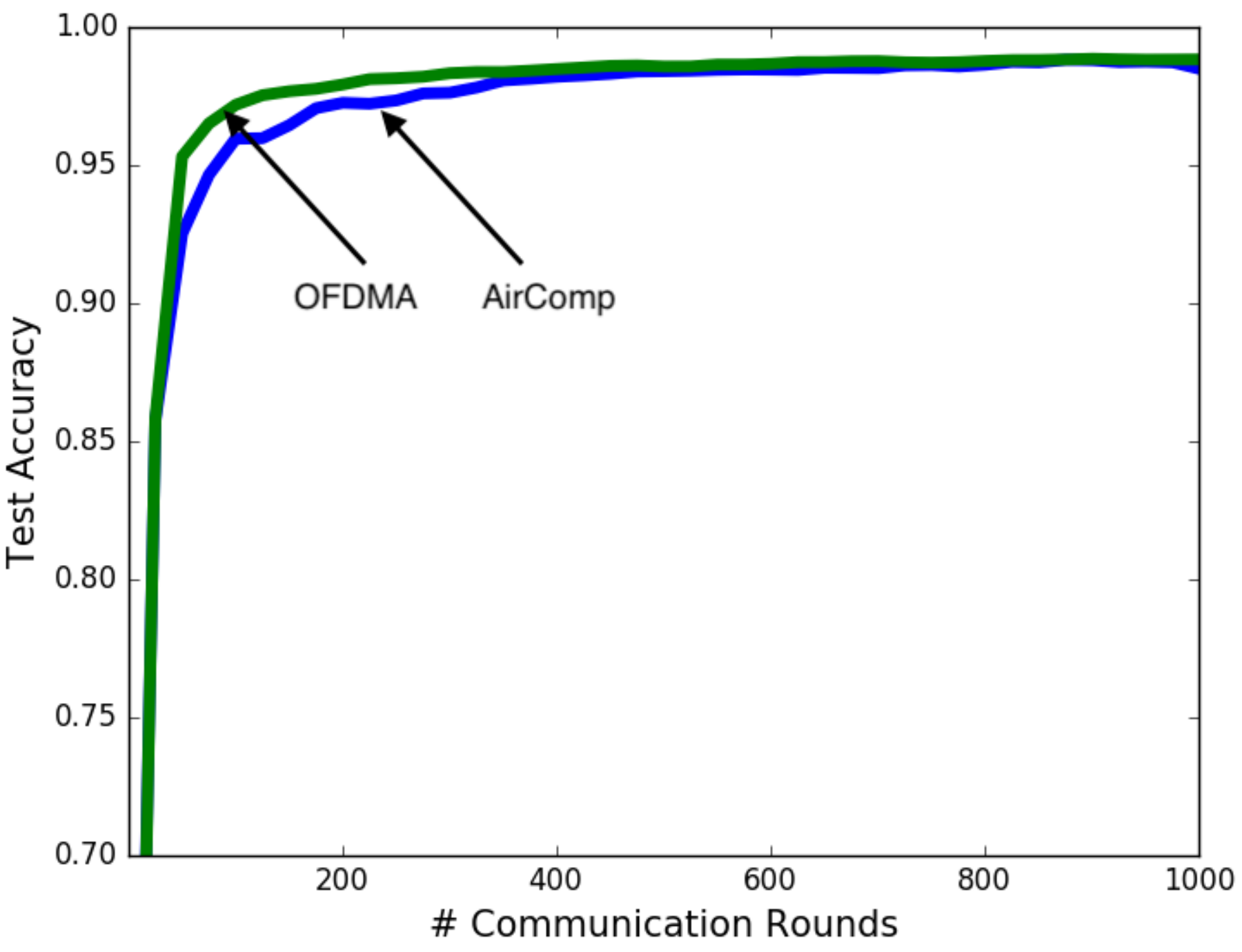}}
\subfigure[Communication latency]{\label{Fig:latency}\includegraphics[width=0.45\textwidth]{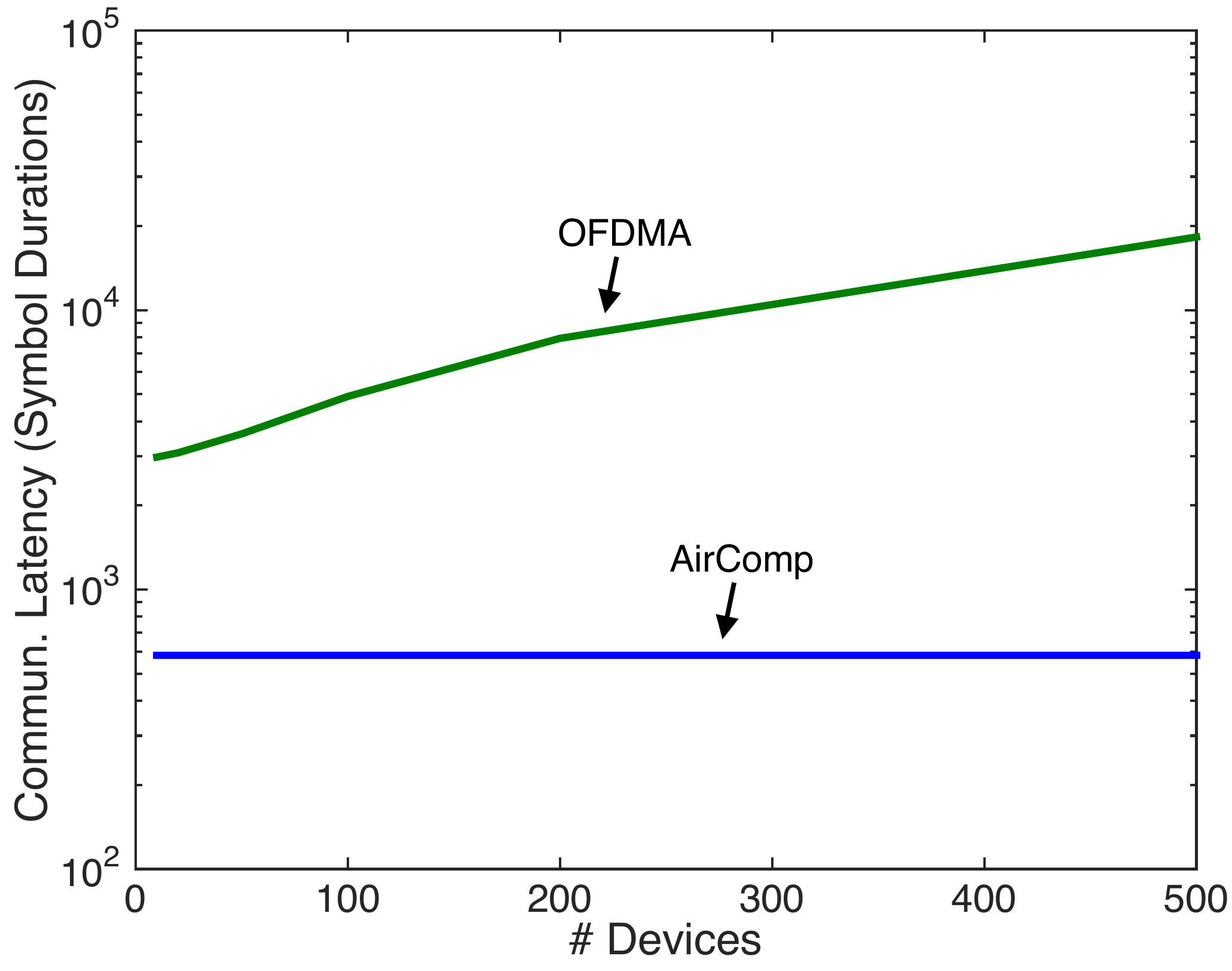}}
\caption{Performance comparison between AirComp and OFDMA in test accuracy (a) and communication latency (b). Consider a federated learning system with one edge server and $100$ edge devices. 
 A 6-layer convolutional neural network (consisting of two $5\times5$ convolution layers with ReLu activation, each followed with $2\times2$ max pooling, a fully connected layer with 512 units and ReLu activation, and a final softmax output layer)
 is trained on the distributed MNIST data for handwritten digit recognition, where the update aggregation is performed by AirComp or OFDMA over a shared broadband channel consisting of $1000$ orthogonal sub-channels. {\color{black}For AirComp, model parameters are analog-modulated and each sub-channel is dedicated for single-parameter transmission; truncated-channel inversion under the transmit-power constraint is used to tackle the channel fading.} For OFDMA, model parameters are first quantized into a bit sequence ($16$-bit per parameter). Then adaptive MQAM modulation is adopted to maximize the spectrum efficiency while maintaining the target bit-error-rate of $10^{-3}$.
 }
 \vspace{2mm}
\label{Fig:performance_comparison}
\vspace{0.5mm}
\end{figure*}

\vspace{-3mm}
\subsection{Distributed Edge Learning}

Driven by gaining low-latency and privacy-aware access to rich mobile data for creating the intelligent IoT, recent years have witnessed the spreading of AI algorithms from the cloud to the network edge, resulting in an active area called \emph{distributed edge learning} \cite{park2019wireless}. As illustrated in Fig. \ref{fig:applications}, a typical algorithm for training an global AI model iterates between two steps : 1) the server receives distributed model updates transmitted by devices over a MAC and applies their average to update the global model; 2) the server broadcasts the updated global model to devices for updating using local data. Step 1) results in a communication bottleneck due to the high dimensionality of each model update (usually comprising millions to billions of parameters) and the multiple-access by many devices. Overcoming the bottleneck is important for alleviating the congestion of the air interface also shared by other types of services and reducing the learning latency for mission critical applications e.g., learning how to deal with ``black-swan'' events in auto-driving. The conventional orthogonal multi-access schemes are inefficient as their underpinning philosophy of interference being a foe causes the required radio resources to scale linearly with the number of devices. To solve the problem intelligently, AirComp has been recently developed as a new air-interface solution for fast model update aggregation in distributed edge learning \cite{yang2020federated,amiri2020machine,zhu2020one}. Thereby, the server directly receives the aggregated version of analog modulated local models/gradients simultaneously transmitted by devices. Compared with the conventional orthogonal multi-access, AirComp can reduce the communication latency by a factor approximately equal to the number of devices without significant loss of the learning accuracy, as shown in Fig. \ref{Fig:performance_comparison}.

Besides the benefit of low multiple-access latency, exploiting AirComp for distributed edge learning has an additional advantage in data privacy enhancement. Note that with advanced model inversion attacks, it is still possible to infer the local training data from the local model updates \cite{liu2020privacy}. As a remedy,  AirComp makes the eavesdroppers can only access to the aggregated updates where each private local one is hidden in the crowd. Moreover, the random perturbation imposed by the channel noise on the aggregated updates is another mask for free that can protect the data privacy.

\vspace{-3mm}

\subsection{Distributed Consensus Control}

Distributed consensus control pertains to a scenario where multiple agents interact with each other with the aim of reaching an agreement over a set of variables of common interest. It is a key operation in a wide range of mission-critical IoT applications such as vehicular platooning and swarm UAV/robot formation control. For example, in vehicular platooning, all the participating vehicles need to reach a consensus on common driving variables of the platoon including its velocity, acceleration, and trajectory. To this end, each agent needs to iteratively update its \emph{information state} (IS), referring the settings of its driving variables, by running an iterative consensus protocol. As illustrated in Fig. \ref{fig:applications}, each iteration of such a protocol comprises 1) a communication step where each agent transmits its IS to other members of the platoon, and 2)  a computation step where an agent updates its IS with the average of those of others. The consensus is reached if all IS's converge to the same value. For mission-critical applications such as vehicle platooning or  swarm UAV, the latency allowed for distributed consensus is ultra-low (e.g., of the order of 1-10 milliseconds). This is essential to ensure the safety of a platoon/swarm usually travelling at a high speed and endow on it the ability of responding to unforeseen events and adapting to complex traffic conditions. The efficient implementation of the consensus protocol using AirComp, which merges the communication and computation steps in each round, can shorten the per-round latency and thereby accelerating the convergence (see e.g., \cite{molinari2018exploiting}). Specifically, in each round, agents simultaneously broadcast their IS’s after analog modulation; as a result, provisioned with full duplexing capability, the agents directly received the average of peers' IS's and updated its own IS with the average. 

It is noteworthy that AirComp for distributed consensus needs to be implemented in a fully decentralized manner without centralized coordination. This is because that, each agent plays a dual role in distributed consensus systems. On one hand, each agent is a client participating in AirComp by contributing its local IS’s to other agents. On the other hand, each agent is also a fusion center that needs to compute a function of the local IS’s from other agents. Designing decentralized AirComp for consensus is a more challenging problem than centralized AirComp for sensing and learning, as it goes beyond the “one-shot” computation of a pre-determined function. Moreover, it needs to account for the interdependence between the multiple functions to be computed at different iterations along the IS update-trajectory for consensus guarantee.

\vspace{-3mm}
\section{Concluding Remarks}
%While 5G is currently being deployed, research on 6G has started aiming at commercializations starting in 2030. 
%The main aim of this article is to introduce a 
%set of new design guidelines to the wireless communication community for accommodating the massive connectivity in 
%the upcoming era of edge intelligence. The introduced learning-driven communication techniques, including multiple access, resource allocation and signal encoding, can break the communication latency bottleneck and lead to fast edge learning.

In the 5G-and-beyond era, we shall witness the widespread deployment of edge computing and AI, network softwarization and virtualization, and massive IoT connecting tens of billions of devices. They jointly constitute a gigantic network that provides ubiquitous computing and intelligence needed by mobile applications and the solutions of large-scale problems faced by our society. As a consequence of the ongoing trend, the focus of 5G-and-beyond wireless networks will shift from communication between human to that between machines.
Due to the ever-increasing device population, the brute-force approach of accommodating massive connectivity with orthogonalized radio resources will soon exhaust the network capacity. 
Motivated by this, in this article, we introduce AirComp, a potential scalable solution for massive IoT, as well as its enabling techniques.
Breaking from the classic design principle that isolates communication from its applications, AirComp explores a new application-specific design approach to boost spectrum efficiency and reduce multiple access latency for massive IoT via seamless integration of sensing, computation, control, and AI. To unleash its full potential in 5G-and-beyond, besides the techniques discussed in this article, advancements in the following directions among others will be also important in our view. 

\begin{itemize}

%\item {\bf New AirComp Protocols}: 
%They include new signalling procedures for initiating the AirComp transmission, new pilot designs for efficient channel estimation, synchronization, and new \emph{hybrid automatic repeat request} (HARQ) protocols customized for AirComp. 

\item {\bf Coexistence Between AirComp and Communications}: It is envisioned that AirComp will coexist with the conventional communication applications in 5G-and-beyond networks.
%The introduction of AirComp results in a new coexistence scenario for 6G with both AirComp and conventional data communications. 
Therefore, new techniques for managing the co-channel interference between the two networks become essential. In particular, due to the simultaneous transmission from multiple  devices, AirComp may introduce more severe uplink interference towards nearby BSs, as compared to the conventional communication users. This thus makes the interference management more difficult.

%\item {\bf AirComp over mmWave and THz}:  The 5G-and-beyond spectrum will include the \emph{millimeter-wave} (mmWave) and \emph{terahertz} (THz) bands. Their drawbacks of severe prorogation loss can be overcome by the deployment of massive MIMO. How to implement AirComp using massive MIMO over the mmWave/THz bands has to be carefully studied, by accounting for the unique channel models for these bands and the efficient hybrid beamforming architecture. 

\item {\bf Large-scale AirComp}: 
5G-and-beyond networks will be highly heterogeneous comprising different types of access points and devices associated with a wide range of services. On the other hand, 5G-and-beyond networks are expected to be 3D by extending the 2D terrestrial networks vertically by adding aerial nodes such as UAVs, balloons, and satellites. To fully exploit enormous data at the edge, developing large-scale AirComp for deployment over such heterogeneous networks is an interesting direction.

\item {\bf AirComp Performance over Practical Networks}:  While theoretical studies show the promise of AirComp, the underpinning ideal assumptions such as perfect CSI may lead to much smaller gains in practice. Therefore, extensive system-level performance evaluation under practical settings are needed for evaluating and refining AirComp before it can be an effective technology in practice.

\end{itemize}

\vspace{-3mm}

\bibliographystyle{IEEEtran}
\bibliography{BibDesk_File}
%
%\vspace{4mm}
%
%\centerline{B\small{IOGRAPHIES}}
%\vspace{2mm}
%
%%\begin{IEEEbiography}
%\noindent{\bf Guangxu Zhu} is a research scientist with Shenzhen Research Institute of Big Data, Shenzhen, China. 
%His research interests include distributed edge learning and 5G/6G systems. 
%%He is a recipient of the Hong Kong Postgraduate Fellowship (HKPF) and a Best Paper Award from WCSP 2013.
%%\end{IEEEbiography}
%%height=1.25in,
%
%\vspace{4mm}
%\noindent{\bf Jie Xu}  is an Associate Professor with the School of Science and Engineering, The Chinese University of Hong Kong (Shenzhen), Shenzhen, China. His research interests include 5G/6G wireless communications.
%
%\vspace{4mm}
%%\begin{IEEEbiography}
%\noindent{\bf Kaibin Huang}  is an Associate Professor with the Dept. of EEE, The University of Hong Kong, Hong Kong. His research interests include mobile edge computing, distributed learning, and 5G/6G systems.
    
    %\end{IEEEbiography}

\end{document}